**Anderson and Stoner Published White Dwarf Mass Limits before Chandrasekhar**

In their engaging recollections of Chandrasekhar's extraordinary career, neither Dyson[1] nor Wali[2] mention that Chandrasekhar was the third person, not the first, to publish a white dwarf mass limit that involved a relativistic treatment of degenerate electrons. It has become a common misconception that Chandrasekhar was the first, so particular clarity on this historical point is warranted.[3,4,5] In fact, Chandrasekhar cited the preceding paper by Stoner,[6] as can be seen in the first paragraph of Chandrasekhar's paper[7] reproduced in the panel on page 46 of Dyson's article[1]. Stoner, in turn, was inspired from a paper by Anderson[8] who also incorporated special relativity to find a white dwarf mass limit. In his book on stellar structure[9], Chandrasekhar mentions Anderson and Stoner on page 422, and in bibliographic note 6 on page 451.

What are the differences between the Anderson[8], Stoner[6], and Chandrasekhar[7] papers? Both Anderson and Stoner found their mass limits for a uniform density star and thus a star whose basic density profile was not derived from a proper treatment of hydrostatic equilibrium. Chandrasekhar subsequently calculated the mass limit using a polytropic equation of state, which did incorporate hydrostatic equilibrium and was known at the time to produce a better approximation to stellar radial density profiles. Although the inclusion of relativity presented in Anderson's paper is perplexing, if not erroneous, Stoner's paper (published in May 1930, before Chandrasekhar's boat trip[10] to England) is lucid. It starts from first principles and culminates with a comparison to observations of presumed white dwarfs of the time. Stoner's mass limit was greater than Chandrasekhar's subsequent polytrope mass limit by about 20%. Incidentally, Stoner also later published a polytropic generalization.[11]

In short, the white dwarf mass limit widely attributed to Chandrasekhar[7] should really be the *specific* mass limit calculated for a polytrope. The insight that a relativistic treatment of degeneracy leads to the existence of *a* mass limit was already identified in the previous papers of Stoner[6] and Anderson[8]. This distinction is commonly blurred, but the correct disambiguation can be found even in Chandrasekhar's own writings.

References
1. F. Dyson, Physics Today **63**, 49 (2010)
2. K. C. Wali, Physics Today **63**, 44 (2010)
3. E.G. Blackman, Nature **440**, 148 (2006)
4. P.-H. Chavanis, Physical Review D **76,** 023004 (2007)
5. M. Nauenberg, Journal for the History of Astronomy, **39**, 297 (2008)
6. E. C. Stoner, *Philos. Mag.* **9,** 944 (1930)
7. S. Chandrasekhar, *Astrophys. J.* **74,** 81 (1931)
8. W. Anderson, *Z. Phys.* **1,** 851 (1929)
9. S. Chandrasekhar, *An introduction to the study of stellar structure,* University of Chicago Press, Chicago (1939)
10. K.C. Wali, *Chandra : a biography of S. Chandrasekhar,* University of Chicago Press, Chicago (1991)
11. E. C. Stoner and E. Tyler, *Philos. Mag.* **11,** 986 (1931)

Eric G. Blackman
(blackman@pas.rochester.edu)
Department of Physics and Astronomy
University of Rochester, Rochester NY, 14618